\documentclass[doublecol]{epl2}

\usepackage[T1]{fontenc}
\usepackage[utf8]{inputenc}
\usepackage{graphicx}
\usepackage{amsmath}
\usepackage{amssymb}
\usepackage{color}
\usepackage{xspace}
\usepackage{multirow}
\usepackage{dcolumn}
\usepackage{comment}

\newcommand{\Ncl}{N_\text{cl}}
\newcommand{\Nrun}{N_\text{run}}
\newcommand{\Trun}{T_\text{run}}
\newcommand{\tmax}{t_\text{max}}
\newcommand{\TTS}{\text{TTS}}
\renewcommand{\eth}{e_\text{th}}

\graphicspath{{./figures/}{./}}

\title{How we are leading a 3-XORSAT challenge: from the energy landscape to the algorithm and its efficient implementation on GPUs}
\shorttitle{How we are leading a 3-XORSAT challenge}

\author{M.~Bernaschi\inst{1} \and M.~Bisson\inst{2} \and M.~Fatica\inst{2} \and E.~Marinari\inst{3,4,5} \and V.~Martin-Mayor\inst{6,7} \and G.~Parisi\inst{3,4,5} \and F.~Ricci-Tersenghi\inst{3,4,5}}
\shortauthor{M.~Bernaschi \etal}

\institute{                    
  \inst{1} Institute of Applied Computing, CNR, I-00185 Rome, Italy\\
  \inst{2} NVIDIA Corporation, Santa Clara, CA 95050, United States of America\\
  \inst{3} Dipartimento di Fisica, Sapienza Universit\`a di Roma, I-00185 Rome, Italy\\
  \inst{4} INFN, Sezione di Roma 1, I-00185 Rome, Italy\\
  \inst{5} CNR-Nanotec, Rome unit, I-00185 Rome, Italy\\
  \inst{6} Departamento de F\'\i{}sica Te\'orica, Universidad Complutense, 28040 Madrid, Spain \\
  \inst{7} Instituto de Biocomputaci\'on y F\'{\i}sica de Sistemas Complejos (BIFI), 50018 Zaragoza, Spain 
}

\abstract{
A recent 3-XORSAT challenge required to minimize a very complex and rough energy function, typical of glassy models with a random first order transition and a golf course like energy landscape.
We present the ideas beyond the quasi-greedy algorithm and its very efficient implementation on GPUs that are allowing us to rank first in such a competition.
We  suggest a better protocol to compare algorithmic performances and we also provide analytical predictions about the exponential growth of the times to find the solution in terms of free-energy barriers.
}

\begin{document}

\maketitle 

\section{Introduction}

On October 2019, we were invited to join a computing challenge on 3-XORSAT problems launched by some colleagues at University of Southern California \cite{USCchallenge}.
The idea behind the challenge was to compare actual performance of the best available computing platforms, including quantum computers, in solving a particularly hard optimization problem.
Quantum computing is becoming practical these days, and many different computing devices based on quantum technologies are becoming available (D-Wave, Google and IBM just to cite the most known).
So it is a natural question to ask, whether any of these quantum devices available today can do better than classical (i.e.\ non-quantum) computing machines.

We decided to join this 3-XORSAT challenge with a proposal combining new algorithmic ideas and a highly optimized GPU implementation.
We are not going to discuss in detail the results of the 3-XORSAT challenge, that will appear elsewhere \cite{USCchallenge}. We just remark that the performances of our algorithm running on commercial Nvidia GPUs are at least 2 orders of magnitude better than those of the other devices that entered the 3-XORSAT challenge: D-Wave quantum annealing processor \cite{DWave}, Memcomputing machine \cite{Memcomputing}, Fujitsu digital annealer \cite{fujitsu} and Toshiba’s simulated bifurcation machine \cite{toshiba}.

This is clearly not the end of the story, as quantum technologies are evolving very fast and presumably will become competitive soon (eventually getting what is called a quantum advantage). Nonetheless, we believe it is very important to clarify what is today the state of art in the ``classical vs.\ quantum computation challenge''.

In the present manuscript we report the ideas and the technical details that make our solving algorithm ranking first as a solver of the hard optimization problems presented at the 3-XORSAT challenge.

The manuscript is organized as follows. First we recap the known physical properties of these hard optimization problems (especially their energy landscape). Then we describe the algorithm we decided to use with a particular emphasis on the use of large number of clones and how this can be implemented efficiently for classical computers. We also provide a description of the technical choices that made our GPU implementation extremely efficient, even if the problem, being defined on a random graph topology, would in principle not be ideal for a platform like GPU. Finally, we discuss the numerical results about the time-to-solution (TTS), proposing an improved way of measuring the largest percentiles of TTS. We finish with a few concluding remarks.

\section{The model and its energy landscape}

The optimization problem that has been presented to the contenders at the 3-XORSAT challenge is the search for the ground state of a model based on Ising variables. The model is well known in statistical physics under the name of \textit{diluted 3-spin ferromagnetic model} \cite{franz2001ferromagnet}. In the computer science literature, it corresponds to a constraint satisfaction problem known under the name of \textit{3-xorsat} \cite{dubois20023}.
In this paper, we use the statistical physics formulation of the model, but switching to its computer science formulation requires just a change of variables.

The model is defined by the Hamiltonian
\begin{equation}\label{eq:Hamiltonian}
  H[\sigma]\equiv -\sum_{(i,j,k)\in E} s_i s_j s_k\;,
\end{equation}
where $s_i=\pm 1$ are $N$ Ising spins. The sum over the set $E$ of triplets $(i,j,k)$ is what defines the interaction topology. The instances provided in the 3-XORSAT challenge were generated on a random regular graph of fixed degree 3. In other words, the set $E$ is made of $N$ triplets randomly chosen under the constraints that in each triplet the 3 indices are different and each index appears exactly in 3 triplets.
From the definition of the Hamiltonian $H[s]$ in Eq.~(\ref{eq:Hamiltonian}), it is clear that the ground state is the configuration $s^*$ with all $s^*_i=1$. However our algorithm, publicly available at \cite{algo}, searches for the ground states without computing the magnetization and having access only to the energy, so there is no need to ``hide'' the solution by a gauge transformation. The organizers of the competition are expected to check that the same is true for all the others contenders.

One may argue that such a model should be easy to optimize, because all interactions are ferromagnetic. However it is well known that such a model shows the same glassy physics of a disordered model \cite{franz2001ferromagnet,ricci2010being} because the 3-spin interaction can be satisfied in many ways and this  generates frustration in the system during the optimization.\footnote{The careful reader may have noticed that the problem of satisfying all interactions in $H[s]$, i.e.\ $s_j s_j s_k = 1$, is equivalent to the problem of solving linear equations modulo 2, $(x_i+x_j+x_k)(\text{mod } 2)=0$ where $s_i=(-1)^{x_i}$. This problem can be solved in polynomial time, e.g.\ by Gaussian elimination. However, as discussed in previous publications \cite{barthel2002hiding}, the problem can be slightly modified preserving the same physical behavior, and making the polynomial algorithm no longer useful. The competition was restricted to algorithms which are robust with respect to such a change: using Gaussian elimination, or algorithms derived from it, was forbidden.} 

Actually, in the 3-XORSAT challenge, an equivalent formulation has been used, where variables are twice in number (one $\eta$ variable is added per each constraint) and variables interact only pairwise \cite{hen2019equation}. This has been done to allow devices implementing only pairwise interactions to enter the competition. The resulting Hamiltonian $H_2[s,\eta]$ is such that  $\sum_\eta H_2[s,\eta]=H[s]$. We have performed such a marginalization on the instances provided, so our algorithm minimizes the cost function $H[s]$ given in Eq.~(\ref{eq:Hamiltonian}).

The 3-spin on 3-regular random graph (3S3R) model offers a paradigmatic example of a \emph{golf course} energy landscape.
The thermodynamics of the model has been exactly solved \cite{mezard2003two,montanari2003nature,krzakala2010following} and its dynamics has been accurately studied numerically \cite{montanari2004cooling,krzakala2010following}.
The picture that comes out from these studies is exactly the one that goes under the name of ``random first order transition'' in the physics of glasses \cite{kirkpatrick1987connections,kirkpatrick1989scaling,castellani2005spin,biroli2012random}.
There exists an exponential (in $N$) number of metastable states that dominate the Gibbs measure below the dynamical critical temperature $T_d\simeq 0.51$, such that for $T<T_d$ ergodicity is broken and the timescale to reach equilibrium diverges with the system size $N$.
The divergence of such a timescale depends on the dynamical algorithm, but for local algorithms, we expect it to be related to the height of some free energy barrier. We discuss this issue in more detail below. For the moment we limit ourselves to observe that the 3S3R model is mean field in nature and thus free energy barriers grow linearly with $N$, implying that the timescales to thermalize and reach the ground state grow exponentially as $\exp(aN)$.

When a local algorithm is run for a time much shorter than $\exp(aN)$, the simulation typically gets stuck at a strictly positive \emph{threshold} energy $\eth$, where the exponential number of metastable states prevents the relaxation dynamics from going deeper in the energy landscape. The value of $\eth$ is not exactly known, but some close bounds are available. Certainly, $\eth$ is below the maximum energy value where metastable states can be found $e_\text{max}=-0.958659$, but we expect it to be also not above the energy value where most of the metastable states are marginal $e_\text{marg}=-0.963594$ \cite{montanari2003nature}. Indeed, marginally stable states correspond to energy minima whose Hessian spectrum has a non-negative support with a lower band edge in zero: that is, they are minima with at least one flat direction. Analytic solutions to the out of equilibrium Langevin dynamics \cite{cugliandolo1993analytical} predicts these are the states reached during the relaxation and recent studies \cite{folena2020rethinking} found that the out of equilibrium dynamics may relax to different energies, but none above $e_\text{marg}$.

The picture to keep in mind is the one of an energy landscape that at the energy level $\eth$ is made of a huge number of marginally stable minima on top of which any local dynamics gets stuck in the search for the few deep wells bringing to lower energies.
As discussed at length in Ref.~\cite{bellitti2021entropic}, the main algorithmic barrier is entropic in nature, that is the relaxation dynamics is not able to proceed further towards the ground state not because the need of jumping over energetic barriers, but because the search needs to find a golf course hole while wandering at the threshold energy.

\section{The quasi-greedy algorithm}

In order to find a set of spins minimizing $H[s]$, one could be tempted to use a greedy algorithm, that is an algorithm accepting only moves decreasing the energy (being interested in fast and local algorithms, we only consider single spin flip moves).
Unfortunately greedy algorithms get stuck at much higher energies than $\eth$, where local minima in $H[s]$ appear \cite{bellitti2021entropic}.

Usually to overcome this problem, one switches on a temperature $T$ and exploits thermal fluctuations to escape from local minima (this is how Simulated Annealing works \cite{kirkpatrick1983optimization}).
However, the 3S3R model does not have a thermodynamic phase transition and for any $T>0$ it is in a paramagnetic phase. This means that in thermal equilibrium, the dynamics will unlikely reach the ferromagnetic ground state in $s^*$.
Even if, by chance, $s^*$ is reached, the dynamics will soon leave that configuration to thermalize again in the paramagnetic state.

In Ref.~\cite{bellitti2021entropic}, a new class of heuristic algorithms has been introduced with the aim of proving that entropic barriers are the main source of computational complexity in the optimization of the 3S3R model. These \emph{quasi-greedy} (QG) algorithms perform with high probability, when this is possible,  a step decreasing the energy, but when they reach a local minimum they keep flipping a spin that enters in at least one violated interaction.
These algorithms have several advantages: (i) they converge fast to the interesting low-energy part of the configurational space; (ii) they keep the system evolving even in presence of many local minima, but without increasing too much the energy (that would make the search ineffective since it would be run in an uninteresting region); (iii) once the ground state $s^*$ is found, the algorithm stops and thus does not escape from the solution of the problem (without the need of checking it after every single spin flip).

Calling $w_k$ the probability of flipping a spin entering $k$ unsatisfied interactions, in Ref.~\cite{bellitti2021entropic} the QG algorithm with $w_0=0$ and $w_2=w_3=1$ was studied numerically. The probability of finding the solution $s^*$ was found to reach  a maximum close to $w_1=0.05$, with a median TTS growing approximately as $\exp(0.0835 N)$.
Starting on these very promising results, we have built here a very optimized version of the QG algorithm.

The QG algorithm can also be viewed as an imperfect Metropolis algorithms not satisfying detailed balance, since by setting $w_0=w_1^3$ we would have an algorithm satisfying detailed balance for the 3S3R model. Setting $w_0=0$ breaks detailed balance, but brings two advantages: large energy jumps are forbidden (they are not strictly required in a search limited for entropic reasons), and once the ground state $s^*$ is found, the algorithm stops.

The latter property is extremely useful because any efficient implementation of the QG algorithm must perform a large number of steps before checking for the energy (that takes a time comparable to the one needed for a sweep of the QG algorithm).
The condition $w_0=0$ ensures that a ground state found during the dynamical evolution will not be lost
between two successive measurements.

\section{The search by rare events requires many clones}

As discussed in \cite{bellitti2021entropic}, the search for the ground state is slowed down by entropic barriers, i.e.\ the search for the right well bringing from the $\eth$ manifold to the ground state configuration $s^*$ is like ``finding a needle in a haystack''.
For this reason, instead of having a single copy of the system evolving for a very long time, it turns out to be more appropriate to follow a large number of copies of the system (evolving independently) for a shorter time, starting each one from a different random initial condition: we call these \emph{clones}.

The rationale beyond this choice is that the evolution on the marginal manifold at $\eth$ is not fast enough to allow a single clone to visit the entire manifold in a reasonable time.
So if the clone starts from an unfavourable initial condition, his search is bound to fail even if it keeps evolving for a very long time.

We are facing a typical phenomenon ruled by rare events: in the large $N$ limit, for a typical initial condition, the QG algorithm gets stuck at $\eth$ and fails to find $s^*$, but there are rare initial conditions that allow the QG algorithm to find the solution $s^*$ in a short time.
The probability of such rare initial conditions (that roughly coincide with the basin of attraction of $s^*$) is exponentially small in $N$, as in any large deviation process.

One has to make a choice between the following two extreme strategies: running a single clone for a time scaling exponentially in $N$ or running a number of clones scaling exponentially with $N$ for a finite time. In principle, one should optimize over all choices in between these extremes, at a fixed total amount of computing time.

Our choice has been to run the largest possible number of clones. This turns out to be the best choice for several reasons. It reduces fluctuations and, if the number of clones is large enough, we can derive analytic predictions for setting the running time to an optimal value (see below). Moreover, it is very unlikely for a single clone (or few clones) to find the solution, while when running a huge number of clones some of them can find the solution, thus allowing us to estimate the mean TTS.
Finally, running a large number of clones is highly beneficial from the coding point of view, since the clones can evolve in parallel, leading to a drastic reduction in running times.

\section{Basic information about the GPU implementation}

We implemented the algorithm described above in a CUDA code that looks for the ground state of the given problem instance using concurrently thousands of clones.
Since spins can only have two values, we use a multi-spin coding technique by packing values from different clones into 32-bit words \cite{jacobs1981multi}. This allows to update 32 distinct clones in parallel by using Boolean operations on the spin words (we use the same random number for the 32 clones in the same word). Moreover we use the natural thread parallelism, evolving different clones in each GPU core. Finally, multiple GPUs can be used simultaneously by executing the same code with different random seeds.
So, in the end, we have three levels of parallelism: $i)$ multi-spin coding; $ii)$ thread level; $iii)$ multiple independent executions on distinct GPUs. Although the code is able to fit the number of threads to the actual number of cores available on the GPU in use, most of the runs have been executed on Volta 100 GPUs featuring 5120 cores. On those GPUs, the total number of clones was $\Ncl=327680$.

One more crucial aspect of our GPU implementation is the partitioning of variables in independent sets. In this way, the spin update procedure can be performed in parallel inside each independent set.

Further details on the GPU implementation, on the optimizations and fine tuning of the code are reported in the Supplementary Material.

\section{Numerical results}

We have been provided with 100 instances for different problem sizes. After some preliminary runs, we decided to focus our attention on sizes 256, 512 and 640, that, once transformed back to the form of a 3S3R model, correspond to $N=128,\,256,\,320$.

\begin{figure}
    \onefigure[width=0.8\columnwidth]{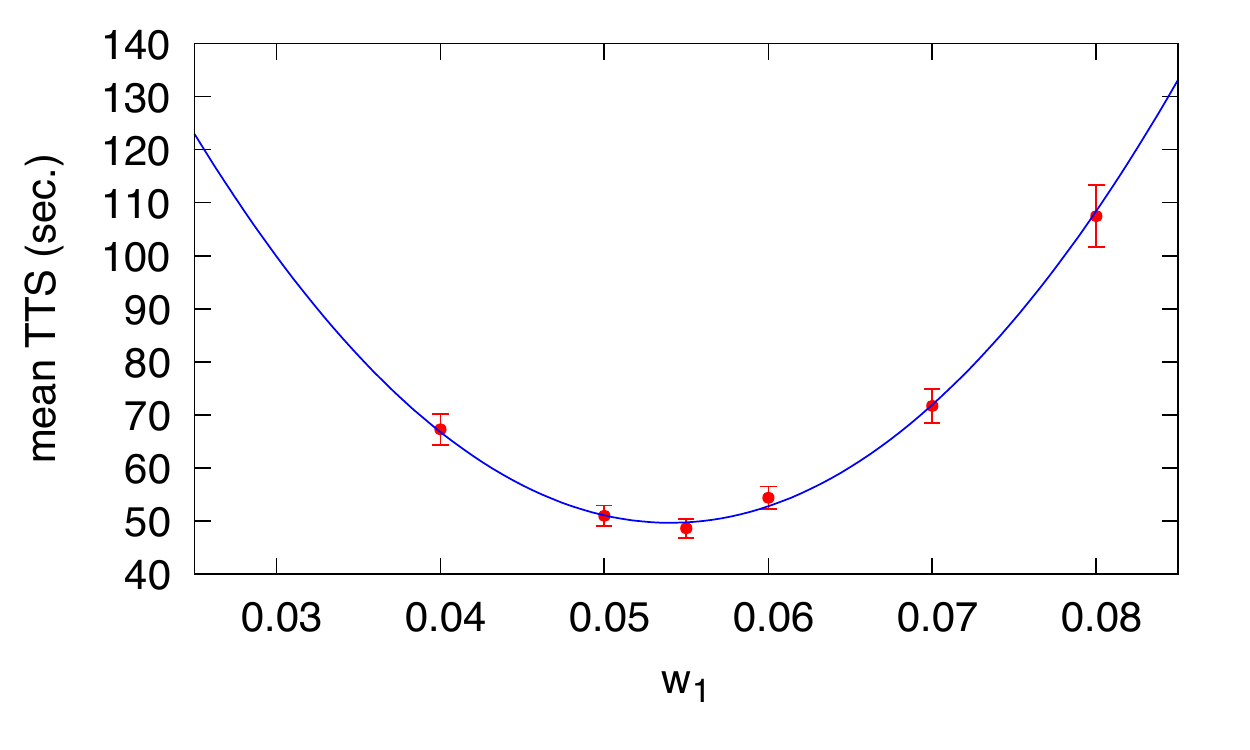}
    \caption{Preliminary runs on instance \#38 of size $N=256$ allowed us to estimate of the optimal value $w_1=0.054(1)$ for the only parameter of the QG algorithm.}
    \label{fig:w1}
\end{figure}

The QG algorithm depends on a single parameter, $w_1$, which is the probability of flipping a variable entering in one unsatisfied interaction and two satisfied interactions. The other parameters are fixed: $w_0=0$ (the ground state is a fixed point of the QG algorithm) and $w_2=w_3=1$ (the QG algorithm decreases the energy whenever possible).
Our preliminary runs also served to optimize over $w_1$. In Fig.~\ref{fig:w1} we show the mean TTS in a given instance of size $N=256$. The quadratic interpolation to the data estimates an optimal value $w_1=0.054(1)$, with a negligible dependence on the problem size (at least for $N\ge 128$). So hereafter we fix $w_1=0.055$.
Although the QG algorithm does not satisfy the detailed balance, we can associate a pseudo-temperature to the value of $w_1$ by the relation $w_1=\exp(-2/\Trun)$, where the latter is the probability of flipping the spin in the Metropolis algorithm running at temperature $\Trun$. It is worth noticing that for $w_1=0.055$ we have $\Trun\simeq 0.69$, which is slightly above the dynamical transition temperature $T_d\simeq 0.51$ \cite{krzakala2010following}.

Being the QG algorithm stochastic, the TTS is a random variable whose probability distribution is often measured via its percentiles $\TTS_p$, defined by $\mathbb{P}[\TTS<\TTS_p] = p/100$.
The organizers of the 3-XORSAT competition asked the participants to estimate the 99-th percentile $\TTS_{99}$ for each of the 100 instances of a given size and to report the median value (over the instances). This is the time required to solve with 99\% probability an instance of median hardness. As we will discuss in detail we are not confident that this is the best metric for evaluating the performance of the proposed algorithms.

\begin{figure}
    \centering
    \includegraphics[width=\columnwidth]{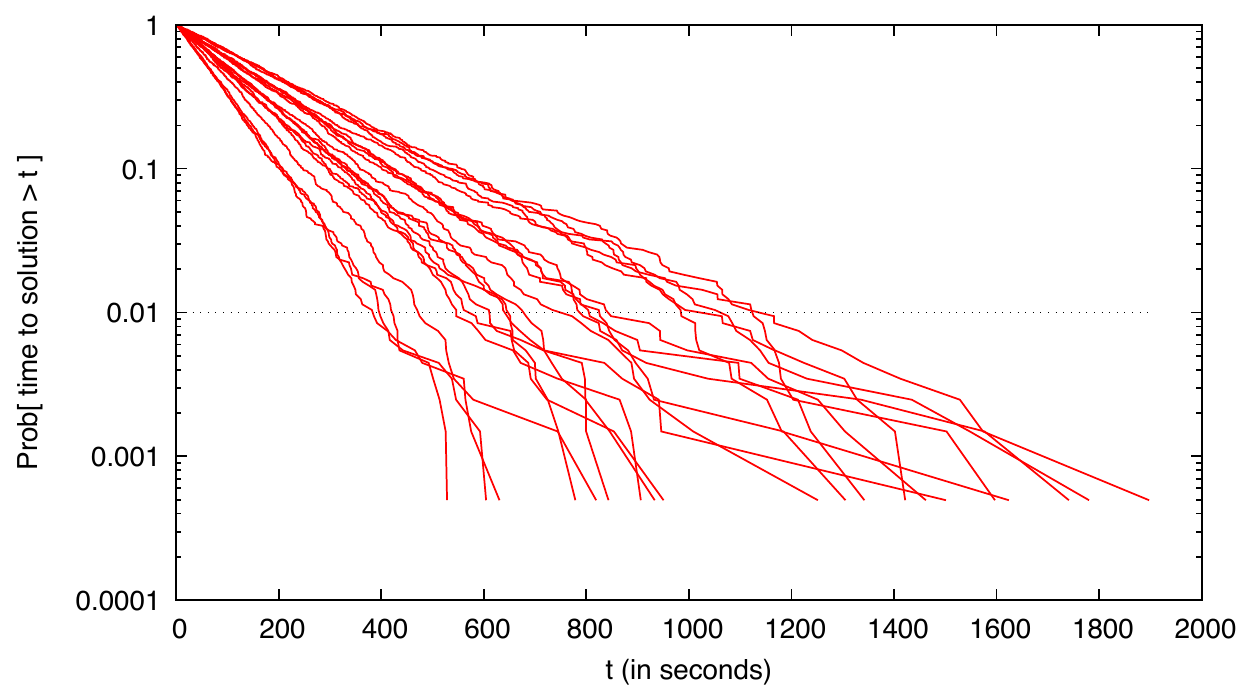}
    \caption{Cumulative probability distribution of TTS for 20 instances of size $N=256$.}
    \label{fig:N256_fig1}
\end{figure}

Most of our simulations have been executed on Nvidia V100 GPUs running in parallel $\Ncl=327680$ clones. 
The TTS for a single run of our QG algorithm is given by the shortest among the $\Ncl$ times each clone requires to reach the solution. Under the assumption, that we have checked numerically with great accuracy, that the cumulative distribution of the single clone time to reach a solution starts linearly in the origin, we have that the TTS (the best of the $\Ncl$ clones) is exponentially distributed as (see SM)
\begin{equation}\label{eq:expTTS}
    \mathbb{P}[\TTS>t] = \exp(-t/\tau)\,.
\end{equation}
A check of the above equation is reported in Fig.~\ref{fig:N256_fig1} where we plot in a semilogarithmic scale the probability that the TTS is larger than a given time $t$ (in seconds) for 20 instances of size $N=256$. Data have been obtained running the QG algorithm 1008 times and sorting the corresponding 1008 values of the TTS. We observe that the exponential distribution (which is linear in a semilogarithmic scale) describes very well the data down to a small probability.
A consequence of this observation is that the TTS of our QG algorithm with a very large number of clones can be perfectly described in terms of the single timescale $\tau$, \emph{the mean TTS}, that depends solely on the particular instance under study.

In Fig.~\ref{fig:N256_fig1} the crossing of the data with the horizontal dotted line determines the value of $\TTS_{99}$. We believe that this is not the best estimate for the time such that the QG algorithm finds the solution with probability 99\%. As a matter of fact, a better estimate, that is affected by much smaller fluctuations, is given by $-\log(0.01)\;\tau$. The latter estimate is much more robust than $\TTS_{99}$ since it is obtained from all the measured TTS values. Moreover, $\TTS_{99}$ requires the execution of the algorithm at least 100 times, whereas $\tau$ can be safely estimated from a much smaller number of measures.

\begin{figure}
    \centering
    \includegraphics[width=\columnwidth]{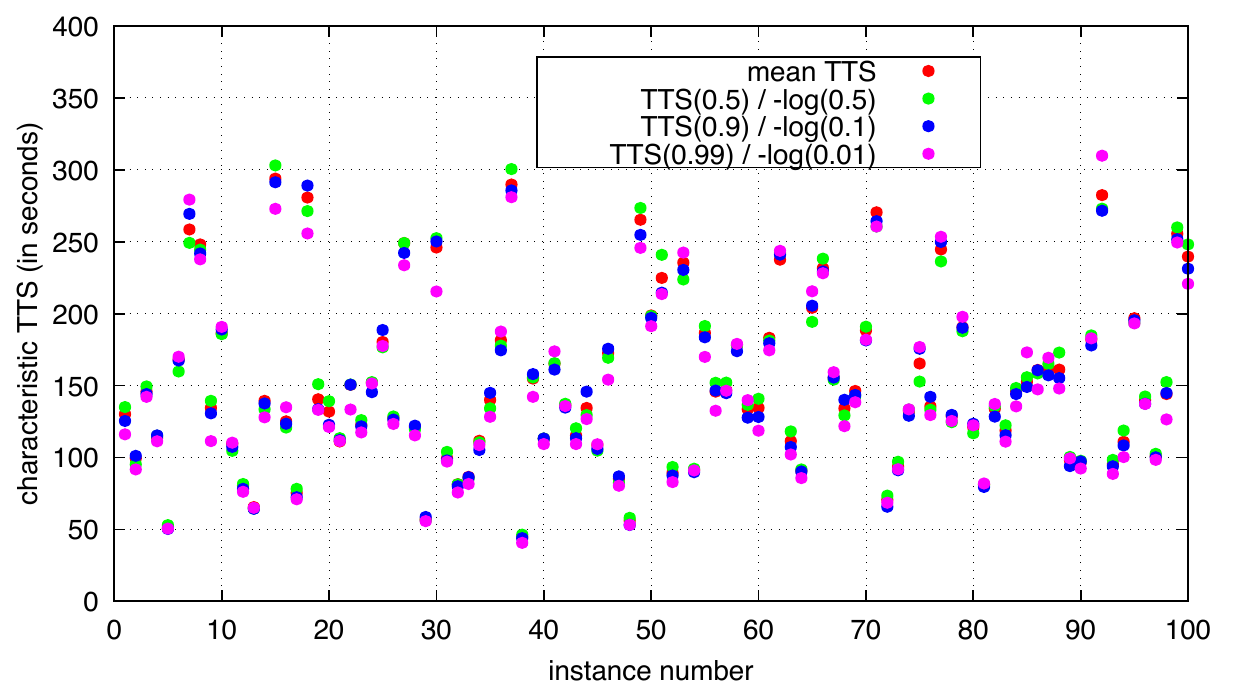}
    \caption{Different estimates of $\tau$, the mean TTS, for the 100 instances of size $N=256$.}
    \label{fig:N256_fig2}
\end{figure}

In general, the value of $\TTS_p$ can be better computed via $-\log(1-p/100)\tau$, after the values of $\tau$ have been estimated. In Fig.~\ref{fig:N256_fig2}, we plot for each of the 100 instances of size $N=256$ the mean TTS $\tau$ and three equivalent estimates obtained from $\TTS_{50}$ (the median), $\TTS_{90}$ and $\TTS_{99}$.
For each instance, the four estimates are very close, whereas they vary a lot when changing the instance. A more careful inspection of the data in Fig.~\ref{fig:N256_fig2} highlights that the mean TTS $\tau$ is always in the middle of the group of the four estimates, while the estimate based on $\TTS_{99}$ is sometimes far from the other.
The above observations suggest that using $\tau$ instead of $\TTS_{99}$ would provide more reliable and stable results in the analysis of algorithms performance.

\section{Running with a short timeout}

What is the best possible way to estimate $\tau$? Obviously, having an unbounded computing power, one could simply execute the search $\Nrun$ times and just take the average of all the TTS values.
But for a process that requires a computing time growing exponentially with the problem size $N$, this naive approach becomes soon unfeasible.

Nonetheless, we deduce from the data shown in Fig.~\ref{fig:N256_fig1} and from the cumulative distribution in Eq.~(\ref{eq:expTTS}) that runs with a very short TTS always exists for any $\tau$, although they become very rare for large values of $\tau$.

So we can adopt a different search strategy. Instead of letting every run to finish reaching the solution $s^*$ (that sooner or later is found, since the dynamical process we simulate is ergodic for finite $N$ values), we can set a timeout $\tmax$ such that the QG algorithm reports a failure if the solution is not found in a time shorter than $\tmax$.

This \emph{early stop} strategy has several advantages.
The use of a timeout prevents very long runs: this is very useful, not only because it stops in advance those unfortunate runs that would take an atypically long time, but also because makes all runs of a similar time duration (and this is very useful when planning a large group of parallel runs).
More importantly, the algorithm with a timeout can also be run for very large sizes, when the algorithm without any timeout would take too long to finish. The data for problems of size $N=320$ have been obtained with this strategy, and a sensible estimate would have been otherwise impossible to get.

\begin{figure}
    \includegraphics[width=\columnwidth]{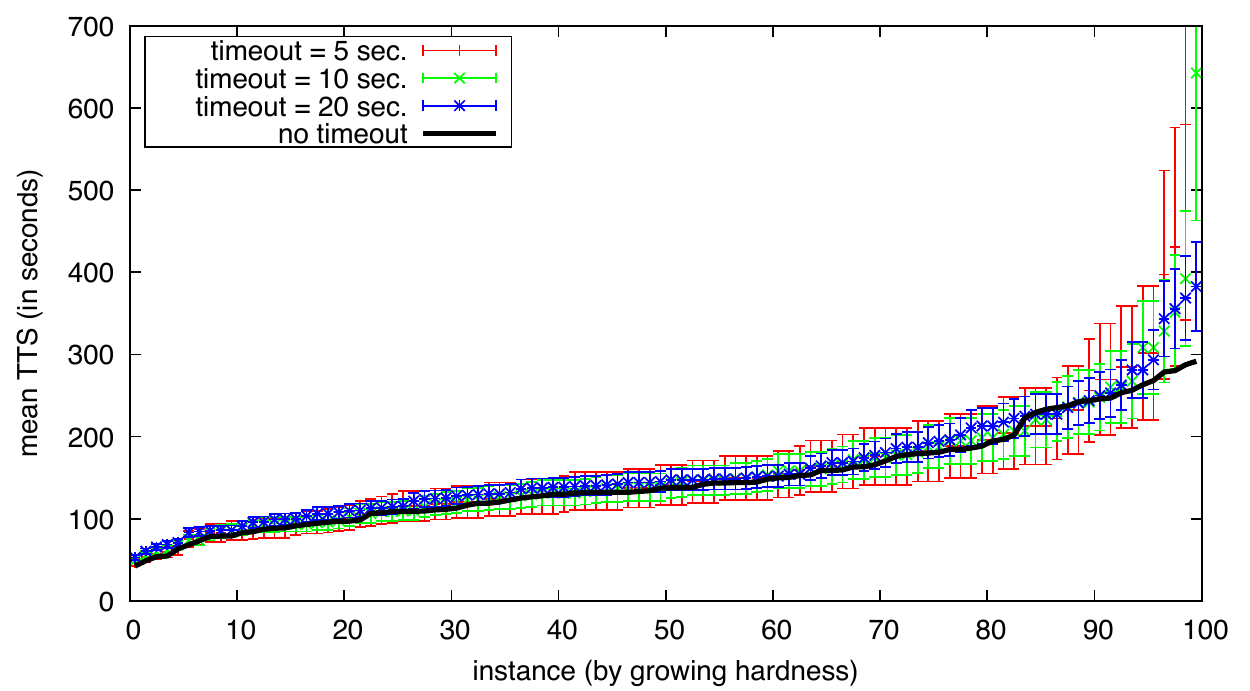}
    \caption{Estimates of mean TTS $\tau$ in all the 100 instances of size $N=256$ obtained from runs with a short timeout.}
    \label{fig3}
\end{figure}

\begin{figure}
    \centering
    \includegraphics[width=0.8\columnwidth]{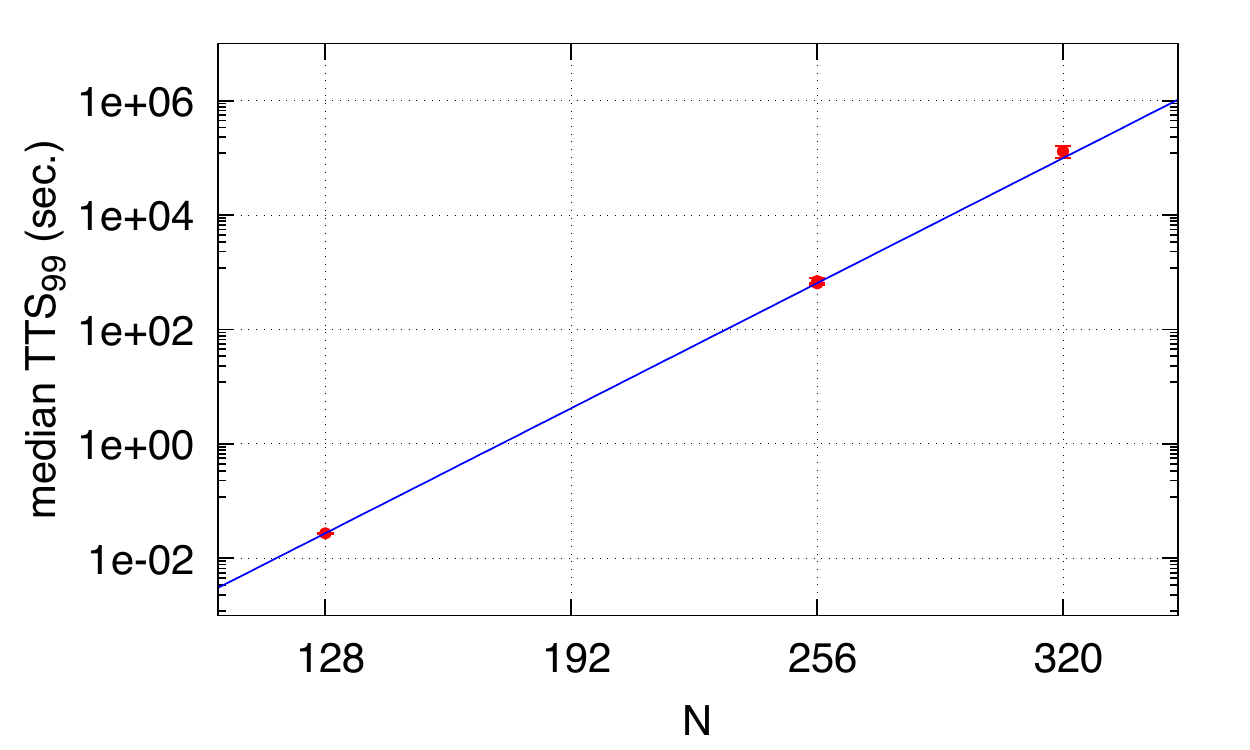}
    \caption{99-th percentiles $\TTS_{99}$ for the median instances of several problem sizes.}
    \label{fig:99perc}
\end{figure}

If the timeout $\tmax$ is much shorter than the mean TTS, $\tmax \ll \tau$, only a small fraction of runs will find the solution. By running $\Nrun$ runs with a timeout $\tmax$, we can estimate $\tau$ from the number $n$ of successful runs as follows.
The posterior distribution on $\tau$ given that we observe $n$ successful runs among $\Nrun$ is proportional to
\begin{equation*}
P(\tau|n) \propto \frac{1}{\tau} \binom{\Nrun}{n} \left(1-e^{-\tmax/\tau}\right)^n \left(e^{-\tmax/\tau}\right)^{\Nrun-n}\;,
\end{equation*}
where the factor $1/\tau$ before the binomial coefficient is the prior on $\tau$ and it is such that, before taking any measurement, the probability measure is uniform on the variable $\ln\tau$.
Since $\tmax\ll\tau$ we can simplify the posterior to the following normalized distribution
\begin{equation*}
P(\tau|n)=\frac{\mathcal{T}_\text{tot}^n}{(n-1)!}\frac{e^{-\mathcal{T}_\text{tot}/\tau}}{\tau^{n+1}}\;,
\end{equation*}
with $\mathcal{T}_\text{tot}=\Nrun\tmax$ being the total running time. Getting an estimate of $\tau$ from this posterior distribution is straightforward and the results are shown in Fig.~\ref{fig3} for timeouts of 5, 10 and 20 seconds.

We see from the data in Fig.~\ref{fig3} that the estimates of $\tau$ from runs with a rather short timeout are very accurate for most of the samples: only for samples whose mean TTS is almost 2 orders of magnitude larger than the timeout did the estimate turn out to be larger, but still compatible within error bars.
In particular we notice that for the median instance the estimates of $\tau$ (and thus of the 99-th percentile $\TTS_{99}$) obtained from runs with a very short timeout are perfectly fine and allow to save a great amount of time.

We summarize in Table \ref{table1} our best estimates for $\TTS_{99}$ in the median instance, and in Fig.~\ref{fig:99perc}, we plot the values reported in the table together with the best fitting exponential growth, $\TTS_{99} \propto \exp(a N)$ with $a=0.0786(4)$.

\begin{table}[!ht]
    \centering
    \begin{tabular}{|llll|} 
    \hline
     & from all runs & from runs & timeout \\
    $N$ & (no timeout) & with timeout & value \\ [0.5ex]
    \hline
    128 & 0.0275 $\pm$ 0.0005 & --- & --- \\
    256 & 640 $\pm$ 20 & 700 $\pm$ 100 & 4.5 \\
    320 & --- & 130k $\pm$ 30k & 450 \\ [1ex] 
    \hline
    \end{tabular}
    \caption{Values of the 99-th percentile $\TTS_{99}$ for the median instances of several sizes. All times are expressed in seconds.}
    \label{table1}
\end{table}

\section{Analytic prediction for exponent $a$ in $\ln\tau\sim a N$}

Although the QG algorithm is heuristic and it does not satisfy the detailed balance condition, we can still obtain an approximate analytical estimate of the exponent $a$ ruling the growth of $\tau$ with $N$, assuming the dynamics takes place in contact with a thermal bath at an effective temperature $\Trun=-2/\log(w_1)$.
In thermal equilibrium, we expect the time to visit the ground state $s^*$ to be related to the free-energy barrier between the paramagnetic state and the ordered state around $s^*$.
We need to compute the free-energy as a function of the magnetization $m=\sum_i s_i / N$.

We consider a $K$-spin model on a $K$-regular random graph (the model we simulated has $K=3$, but is worth presenting analytic computations for a generic $K$ value). In order to set the magnetization to an arbitrary value, we add an external field $b$ to the Hamiltonian: $H[s]-b \sum_i s_i$. Using the cavity method for sparse models \cite{mezard2001bethe,mezard2003cavity,mezard2009information} we can write the free-energy at temperature $T=1/\beta$ in the following variational form
\begin{eqnarray}
    -\beta f &=& \log(Z_i) + \log(Z_a) + K \log(Z_{ai}) - b\,m\;,\label{eq:f}\\
    Z_i &=& \frac{2 \cosh(\beta(Ku+b))}{(2 \cosh(\beta u))^K}\;,\nonumber\\
    Z_a &=& \cosh(\beta) \left(1+\tanh(\beta) \tanh(\beta h)^K\right)\;,\nonumber\\
    Z_{ai} &=& \frac{\cosh(\beta(u+h))}{2\cosh(\beta u)\cosh(\beta h)}\;,\nonumber
\end{eqnarray}
that needs to be extremized with respect to the external field $b$ and the cavity fields $u$ and $h$. The saddle point equations read
\begin{eqnarray}
    m &=& \tanh(\beta(Ku+b))\;,\nonumber\\
    h &=& (K-1)u+b\;,\label{eq:spe}\\
    \tanh(\beta u) &=& \tanh(\beta) \tanh(\beta h)^{K-1}\;.\nonumber
\end{eqnarray}
The paramagnetic solution to Eq.~(\ref{eq:spe}) has $u=h=m=0$ and $f=f_\text{para}\equiv-\log(2\cosh(\beta))/\beta$.
The ferromagnetic solution has $u,h,m>0$ and it exists only for temperatures $T<T_\text{sp}$, with the spinodal temperature given by
\begin{equation*}
    T_\text{sp}^{-1} = \min_x \left\{\text{atanh}\left[ \tanh(x) \tanh\big((K-1)x\big)^{K-1} \right] \right\}\;.
\end{equation*}
For $K=3$, we have $T_\text{sp}=0.980548$, and below this value, we can compute the barrier separating the paramagnetic and the ferromagnetic states.

\begin{figure}
    \centering
    \includegraphics[width=\columnwidth]{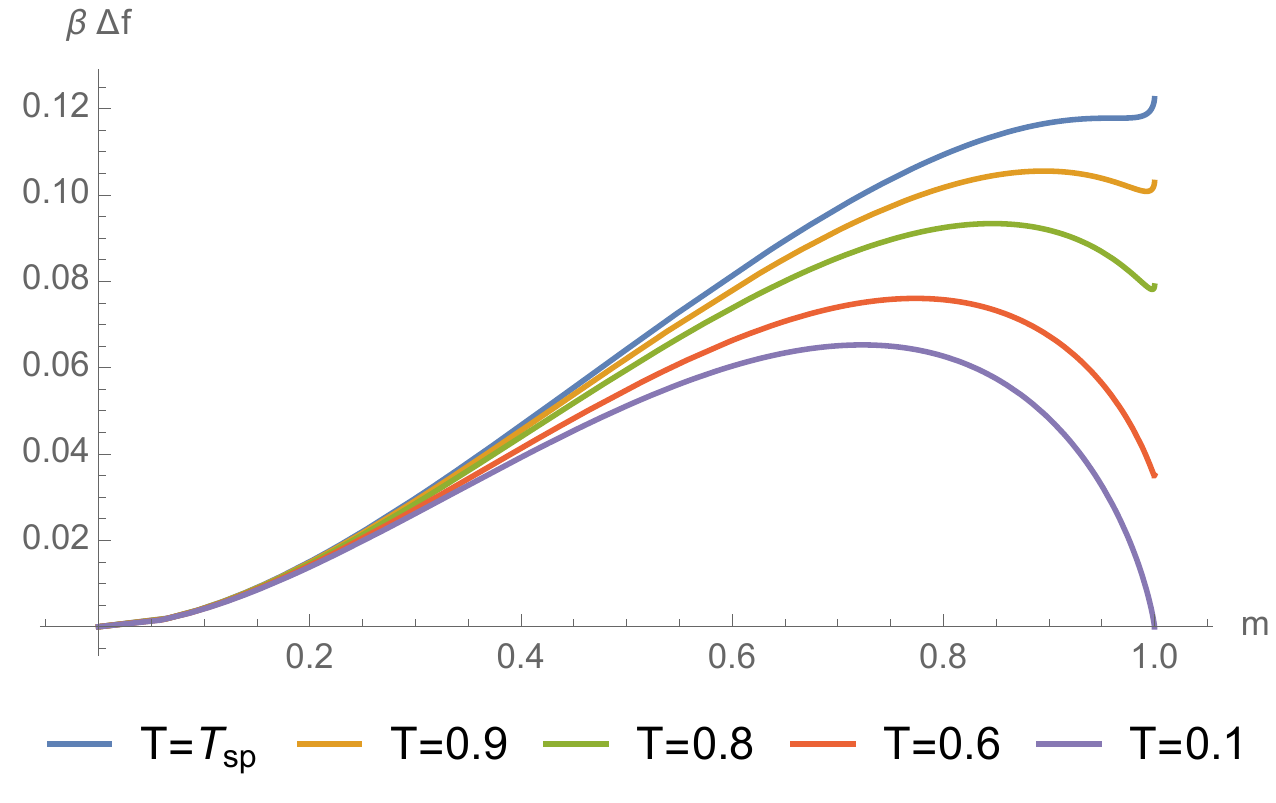}
    \caption{The free-energy as a function of the magnetization at different temperatures below the spinodal one.}
    \label{fig:fm}
\end{figure}

In Fig.~\ref{fig:fm}, we plot $\beta \Delta f=\beta(f-f_\text{para})$ as a function of the magnetization $m$ for several temperatures below the spinodal one. We notice that the ferromagnetic state has always a higher free-energy than the paramagnetic state. Indeed this model has no thermodynamic phase transition, but only a dynamical transition at $T_d$.
Although the ferromagnetic state never dominates the Gibbs measure in the large $N$ limit, we can estimate the time for reaching it by thermal fluctuations via
\begin{equation*}
    \tau \approx \tau_0 \exp\left[N \beta \max_m \Delta f(m)\right]\;,
\end{equation*}
where $\tau_0$ is a microscopic timescale that we expect to depend mainly on the temperature and to diverge in the limit $T\to T_d^+$.

\begin{figure}
    \centering
    \includegraphics[width=\columnwidth]{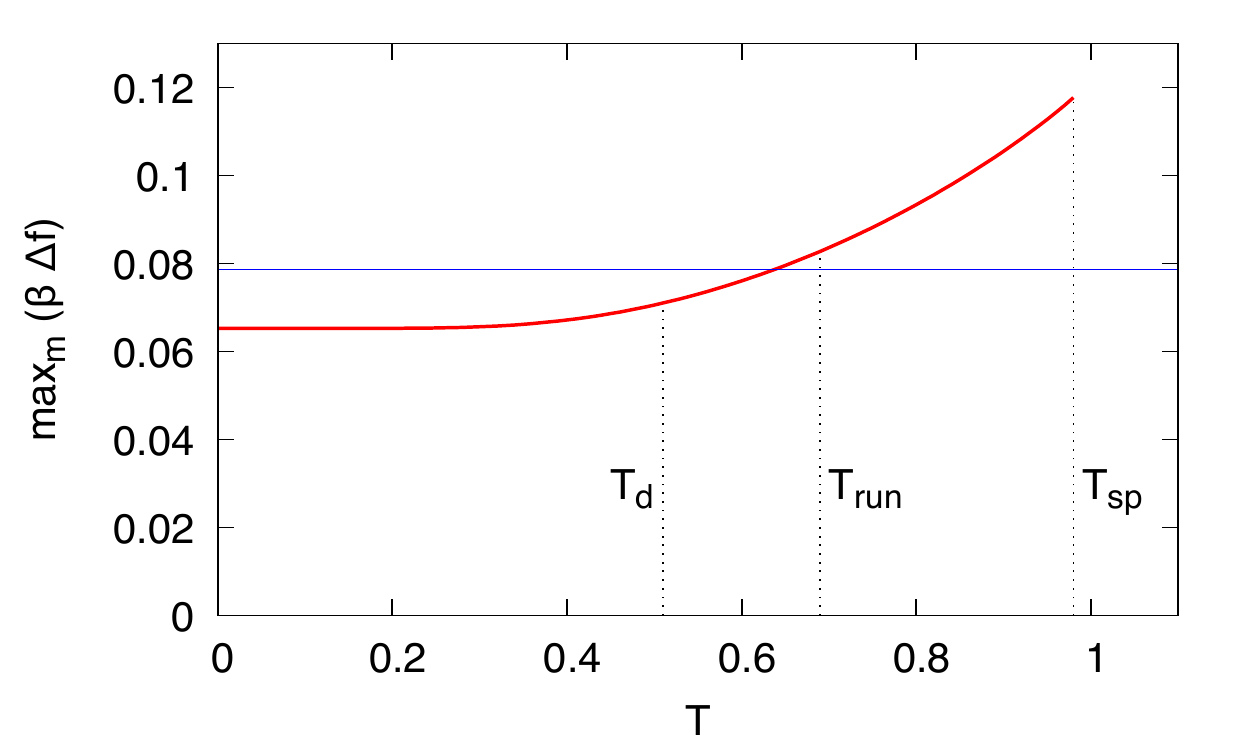}
    \caption{The free-energy barrier that determines the growing of the timescale to visit the ground state by thermal fluctuations.}
    \label{fig:barriera}
\end{figure}

In Fig.~\ref{fig:barriera}, we show the free energy barrier as a function of the temperature, together with the relevant temperatures discussed so far. The horizontal line corresponds to the actual growing rate of the QG algorithm.

Several observations are in order. The analytic prediction based on the thermodynamic free-energy barrier $\beta \max_m \Delta f(\Trun) = 0.0827249$ is a very good approximation to the actual rate.
According to data shown in Fig.~\ref{fig:barriera}, in principle a purely thermal algorithm could be even faster, but the divergence of $\tau_0$ approaching $T_d$ implies the existence of an optimal running temperature slightly above $T_d$. We believe $\Trun$ used in our simulations is close to such an optimal temperature.

\section{Conclusions}

We have described the algorithm and the strategy that are allowing us to rank first and well above all the other contenders in the 3-XORSAT competition that asked to solve a very hard optimization problem.

Although finding the ground state to the 3S3R model requires a time growing exponentially with the system size, we have discussed the strategy that allowed us to reduce the growing rate to a rather small value.

The detailed analysis of the running times that we have carried on suggests that the dogma of measuring $\TTS_{99}$ for the median problem is not the best measure of algorithm performance, as it requires enormous computing resources without any clear advantage with respect to measuring the median TTS. The strategies discussed above (cloning and restarting) avoid the longest runs and make $\TTS_{99}$ of limited practical relevance. We propose to rank algorithms according to the median TTS to solve the hardest instances, given that fluctuations among instances are far more severe.

The successful connection between the dynamical behavior of the QG algorithm and the thermodynamic barrier of the 3S3R model suggests that the statistical physics description of algorithms in terms of the energy landscape is a key tool towards their full comprehension and the proposal of possibly better optimization algorithms \cite{zhou2020circumventing}.

\acknowledgments
We are grateful to Itay Hen for the helpful discussions.
This work was supported by MINECO (Spain) through Grant No.~PGC2018-094684-B-C21 (partially funded by FEDER) and the European Research Council under the European Union's Horizon 2020 research and innovation program (Grant No.~694925-Lotglassy).
This research used resources of the Oak Ridge Leadership Computing Facility at the Oak Ridge National Laboratory, which is supported by the Office of Science of the U.S. Department of Energy under Contract No.~DE-AC05-00OR22725.

\bibliographystyle{eplbib}
\bibliography{biblio}

\begin{widetext}
\begin{center}
\textbf{\Large Supplementary Material for manuscript\\
``How we are leading a 3-XORSAT challenge:\\
from the energy landscape to the algorithm\\[4pt]
and its efficient implementation on GPUs''}
\end{center}
\end{widetext}

\section{Derivation of the exponential distribution for the TTS}

We derive here a central-limit-theorem like result that backs Eq.~(2) in the main text.

Let $p_1(t)$ be the probability that a single clone has not found the ground state after running for time $t$. Let us further assume that $p_1(t)$ can be expanded as
\begin{equation}
p_1(t)=1-\frac{t}{t_0} +a_2 t^2+\ldots\,.
\end{equation}
Let us now assume that we run $\Ncl$ clones in parallel in our device. The probability that none of them hits the ground state after running for time $t$ is
\begin{equation}
p(t)= [p_1(t)]^{\Ncl}\,.
\end{equation}
We shall now take the limit of a large number of clones by keeping fixed $\hat t= t N_\mathrm{cl}$
\begin{equation}
\log p(t=\hat t/N_\mathrm{cl}) = N_\mathrm{cl}\log p_1(t)= \hat t/t_0 + {\cal O}(1/N_\mathrm{cl})\,,
\end{equation}
which tells us that the limiting law is, indeed, exponential. Furthermore, a comparison with Eq.(2) in the main text tells us that $\tau= t_0/N_{\mathrm{cl}}$.

\section{Details on the GPU implementation}

The initial version of our code used a single CUDA thread to update
the spins in each multi-spin coded replica. In this version, each of the thousands of threads running on the GPU processed the graph one
node at a time by reading from global memory its $32$ spin values and those of
its neighbors, executing the spin update procedure and, if necessary, updating
the spin word of the source node by writing it back to global memory.
The information about the connectivity of the graph (that is the same for all the simulated replicas) was stored in the fast (cache-like) {\em shared memory} of the GPU. 
While this
initial approach already delivered a good performance, we eventually implemented a number of
optimizations that allowed the final version of the code to speedup considerably. First of all, while the GPU can run large numbers of threads
simultaneously, using a thread {\em per} msc-replica\footnote{Hereafter we refer to each group of 32 independent multi spin coded clones as a ``msc-replica''}  made the processing of each of them a
serial procedure (if we ignore the parallelism due to the multi-spin coding). Our first step of optimization consisted in exposing
parallelism in the spin update of each msc-replica in order to process multiple nodes in
parallel by multiple threads, speeding up considerably the update process.
Also, using one thread per msc-replica made it necessary to store the msc-replica data in
global memory as the limited size of the faster, on-chip shared memory was insufficient to hold a distinct msc-replica for each thread in a block
(one can have up to 1024 threads in a single block). This
represented one of the main bottlenecks of the code performances since the arithmetic density
of the kernel is not high enough to hide the latency of the global memory
accesses. However, using multiple threads {\em per} msc-replica, the number of
replicas processed per block shrinks, making possible to move all the data into
the shared memory. Moreover, since to process each node a thread must read its
whole adjacency list from memory, we employed a low level optimization to allow
threads to load an entire 6-element adjacency list with a single load
instruction, achieving higher bandwidth in reading data from the shared
memory.

Our first optimization consisted in an algorithmic improvement aimed at exposing
parallelism in the spin update procedure. Since nodes that are not directly
connected can be processed in parallel,
before starting the simulations, we pre-process the interaction graph that determines the instance of the problem by partitioning it into
independent sets. In this way, the spin update procedure can be performed in
parallel inside each set (each one represents an {\em anti-clique} in the graph jargon). We process each msc-replica (and we recall that they are multi-spin coded) with a warp (that is a group of 32 CUDA threads). Each warp progresses in its search for the ground state by processing the graph one
independent component at time, and for each component by updating the spin values in parallel. A warp synchronization is performed after the processing of each component in order to avoid race conditions.

In order to partition the graph in independent sets, we employ a simple greedy strategy in which the graph is scanned multiple times, each time trying to build a set of independent nodes of maximum size up to a fixed value, until all the nodes have been assigned to a set. Since we process each replica by using a warp, we expected that the optimal size of a set had to be a multiple of $32$, in order to keep the processing of components as balanced as possible among the warps' threads.  Preliminary tests showed that the best performance was obtained by partitioning the graphs in sets of $32$ nodes.

The choice of a simple partitioning scheme is motivated not only by the limited size of the graph that has, at most, few hundreds nodes, but also by the fact that the graph does not change during the computation. 
The partitioning is done just once in the beginning of the computation and takes a negligible fraction of the total execution time.

After the graph partitioning, the initial spin values for each replica are
generated randomly using the Parisi-Rapuano pseudo random number generator [Physics Letters B, \textbf{157} (1985) 301]
Since each node is part of exactly three triangles (each triangle represents a clause of the 3-xorsat formula) the graph is represented by a $N\times 6$ matrix, where $N$ is the number of nodes. 

The graph partitioning is represented with two arrays: one that contains a permutation of the node indices such that all the nodes belonging to the same independent set are contiguous and
the second, with an element per set, that contains the size of each set (maximum $32$). The samples are stored as columns of a $N\times NumSamples$ matrix, where $NumSamples$ (the number of msc-replicas) is specified as input to the code (a typical choice is twice the number of computing cores available on the GPU).

We use the smallest integer type for the graph and permutation buffers,
sufficient to represent the node indices. For graphs with up to $256$ or fewer nodes,  we
use the \verb|unsigned char| type and for larger graphs we used \verb|unsigned short|
(this imposes a limit of $65536$ nodes that, however, is well beyond any chance of being solved for a long time).  For the
sample matrix, we used 4-bytes integers, each one representing 32 spins. 

The kernel in charge of the spin update is launched with one warp per sample. The kernel is divided in three logical parts. Initially, the warps copy from global to shared memory the
input data (spin values, graph adjacency matrix and partitioning), then each warp executes the spin update procedure on its samples
until the ground state is found or a maximum number of attempts has been made, and finally the updated spins are copied back to the permanent buffers in global memory.
There are multiple attempts to update the spins within a single execution of the kernel to dampen the overhead of invoking it.
A good trade-off is to {\em sweep} the whole system one hundred times at
each kernel execution. When a warp finds that its system has reached the ground state, the execution continues until the requested number of attempts has been completed but the spins are no longer updated. This choice simplifies the control flow since there is no need to notify the other threads of the termination (communication among threads of different blocks is, somehow, discouraged by CUDA programming best practices). However, the exact iteration in which the system has reached the ground state is saved and reported at the end of the execution. 

Each warp accesses the adjacency lists of the whole graph, the partitioning
permutation and one column of the sample matrix. The graph and permutation data
are read-only and are the same for every warp in the grid so we use a {\em single}
shared buffer for each block. For the samples, on the other hand, we use a
separate array for each warp in the blocks.

The limited size of the graphs combined with using a single warp per sample,
allows us to fit all the data required to update the spins in shared memory making the kernel much faster by
drastically limiting the latency of the memory accesses.
For example, for the case $N=256$ using blocks of eight warps ($256$ threads), each
block requires $10,752$ bytes of shared memory, limiting the number of resident
blocks on each SM to $6$ ($1,536$ threads/SM), which means that $49,152$ samples
are processed per SM concurrently (taking into account the factor 32 due to the multi-spin coding).

In addition to the warp-level parallelization of the spin update procedure and to moving
all the graph and spin buffers to the shared memory, 
we also employed a further, low level optimization aimed at reducing the number 
of memory load operations necessary for reading the adjacency lists from shared
memory. Since each list contains exactly $6$ vertices, at most $6$ loads are
necessary to read a list into the registers. Lowering this number allows to 
reduce further the number of instructions executed to read the same amount of data,
resulting in a reduced latency of the accesses in shared memory. To that purpose, we
increased the size of the lists by adding two unused elements. This makes the
graph buffer $30\%$ larger but it also makes the adjacency lists size exactly equal to either $8$ or $16$ bytes, depending on whether the nodes are represented with char
or short int. It is then sufficient to wrap the 8-elements arrays representing
the lists in either a \verb|union| or a \verb|struct| and reading them from
memory as a single piece of data in order for the compiler to use a single 64-
or 128-bit load for each one (LDS.U.\{64,128\}). Finally for the random numbers required by the update procedure, we resort to the generator provided by the $CURAND$ library, part of the CUDA programming environment.

\end{document}